\pdfoutput=1

\documentclass[twocolumn]{aastex61}

\usepackage{savesym}
\savesymbol{tablenum}
\usepackage[separate-uncertainty=true,multi-part-units=single]{siunitx}
\DeclareSIUnit\year{yr}
\restoresymbol{SIX}{tablenum}
\usepackage{amssymb}
\usepackage{amsmath}

\shorttitle{Bayesian Inference of Reaction Rates in Icy Mantles}
\shortauthors{Holdship, Jeffrey, Makrymallis, Viti \& Yates}

\begin{document}

\title{Bayesian Inference of the Rates of Surface Reactions in Icy Mantles}

\author[0000-0003-4025-1552]{J. Holdship}
\affil{Department of Physics and Astronomy, University College London, Gower Street, WC1E 6BT}
\email{jrh@star.ucl.ac.uk}

\author{N. Jeffrey}
\affil{Department of Physics and Astronomy, University College London, Gower Street, WC1E 6BT}

\author{A. Makrymallis}
\affil{Department of Physics and Astronomy, University College London, Gower Street, WC1E 6BT}

\author{S. Viti}
\affil{Department of Physics and Astronomy, University College London, Gower Street, WC1E 6BT}

\author{J. Yates}
\affil{Department of Physics and Astronomy, University College London, Gower Street, WC1E 6BT}

\begin{abstract}
Grain surface chemistry and its treatment in gas-grain chemical models is an area of large uncertainty. Whilst laboratory experiments are making progress, there is still much that is unknown about grain surface chemistry. Further, the results and parameters produced by experiment are often not easily translated to the rate equation approach most commonly used in astrochemical modelling. It is possible that statistical methods can reduce the uncertainty in grain surface chemical networks. In this work, a simple model of grain surface chemistry in a molecular cloud is developed and a Bayesian inference of the reactions rates is performed through MCMC sampling. Using observational data of the solid state abundances of major chemical species in molecular clouds, the posterior distributions for the rates of seven reactions producing CO, CO$_2$, CH$_3$OH and H$_2$O are calculated, in a form that is suitable for rate equation models. This represents a vital first step in the development of a method to infer reaction rates from observations of chemical abundances in astrophysical environments.
\end{abstract}

\keywords{Physical Data and Processes: astrochemistry--- ISM: molecules--- ISM: dust --- methods: statistical--- methods: numerical}

\section{Introduction}
Dust grain chemistry plays an important role in the physical processes happening deep inside dark molecular clouds during star formation \citep{Draine2003,Williams2015}. These dust grains are vital to every part of the star formation process and ultimately contribute to the basic matter from which icy planetesimals are formed \citep{Vandishoeck2004}. It is in fact evident that molecules such as water and methanol in dust grain ice mantles are primarily formed through solid state chemistry rather than accreted directly from the gas phase \citep{Parise2005,Ceccarelli2007}. In recent years, even more complex molecules have been observed in both prestellar cores and star forming regions \citep[see reviews by][]{Herbst2009,Caselli2012}, some of which can not currently be explained by pure gas phase chemistry. Therefore, chemical reactions leading to simple as well as complex molecules must occur on the surface of icy dust grains. \par
Experimentally, it has been known since the work of \cite{Hagen1979} and \cite{Pirronello1982} that grains can be chemical nanofactories on which surface reactions, UV photons and cosmic rays radiation can synthesize complex molecules and even prebiotic species, starting from simple atoms or molecules such as H, C, O, N, CO. Therefore, understanding the surface chemistry that takes place on dust grains is key to understanding not only the origins of stars, but also how rocky and gaseous planets are formed.\par
Initially, surface reaction networks in chemical models were based on chemical intuition and gas phase chemistry analogues. However, over the past two decades, laboratory astrochemists have been using experimental techniques to test and evaluate surface reactions. As a result, the efficiencies of reaction routes are being properly explored and important information on how molecules form on grain surfaces is being revealed \citep[see][for a review]{Williams2015}. The first experimental work on the dust surfaces studied the formation of molecular hydrogen \citep{Pirronello1997}. Several more experiments followed studying either the formation of more complex molecules \citep[e.g.][]{Watanabe2005,Ioppolo2009,Minissale2015} or the ice morphology and ice mantle mechanisms \citep[e.g.][]{Fraser2004,Collings2006}. Surface reactions can be experimentally investigated within a constrained range of laboratory conditions. Typically, these conditions include different atomic fluxes, ice temperatures, ice morphologies, and mixture ratios, as well different energetic processes. The aims of the experiments are to investigate surface molecule formation, desorption and diffusion. However, experimental data for interstellar ices is limited, the main reason being that the experimentation process is neither simple nor fast. In order to make the best use of experimental resources, the chemical data that models require needs to be prioritized according to what will have most impact.\par 
Bayesian methods are widely used in astronomy as a means of deriving posterior probability distributions for model parameters from observations \citep[eg.][]{Palau2014,Schmalzl2014,Bevan2018}. It is the de facto standard in the field of cosmology but is becoming more and more widely used in other areas of astronomical research \citep[eg.][]{Lomax2016,Testi2016}. In astrochemistry, Bayesian inference has been used to derive parameters such as the gas density and cosmic ray ionization rates within a dark molecular cloud from observations of species in the gas and ices using chemical models \citep{Makrymallis2014}.\par
In this work, we investigate the chemistry itself, studying the rates of reactions on the dust grain surfaces in an attempt to infer their rates and provide a list of reactions for which an accurate rate is particularly important. This is the first such work in an astrochemical context but Bayesian methods have been used to determine rate coefficients for combustion chemistry on Earth \citep{Prager2013}. This work represents a necessary first step in which we determine whether we can use a reduced chemical model and very simple observational constraints to learn more about the parameters in a grain surface chemical model.\par
The grain chemistry model used in this work is described in Section~\ref{sec:model}. The inference process including the choice of MCMC sampler is presented in Section~\ref{sec:inference}. The results of our analysis are presented in Section~\ref{sec:results} along with additional discussion in Section~\ref{sec:discussion}. Finally, our conclusions are discussed in Section~\ref{sec:conclusions}.
\section{The Chemical Model}
\label{sec:model}
A simple chemical model was developed that considers only the solid state chemistry in the ice mantles of dust grains in a dark molecular cloud. The simplified model is a time-dependent single-point model that generates a time series of solid phase molecular abundances as a function of the physical conditions of the molecular cloud and the chemical parameters of the defined chemical network. The chemical network consists of 23 species and 24 surface reactions that are listed in Table~\ref{table:species} and Table~\ref{table:reacs} respectively.\par
To model the surface chemistry of a dark cloud the abundance of each solid species is derived by solving rate equations for grain-surface chemistry. The formation and destruction mechanisms for a species $i$ are given by the following kinetic equation:
\begin{equation}
\frac{dn_i}{dt} = \sum_{l,m}k_{lm}^in_ln_m - n_i\sum_{i \not= r}k_rn_r -k_i^{des}n_i + k_i^{ads}n_{i,gas},
\end{equation}
where $k_{lm}^i$ is the reaction rate of all the reactions between species $l$ and $m$ that produce $i$, $n_i$ is the concentration of species $i$ (with the subscript gas indicating the concentration of the species in the gas-phase), $k_r$ represents the reaction rates of all the reactions where species $i$ participates as reactant, while $k_i^{des}$ and $k_i^{ads}$ are the desorption and adsorption rates.\par
The reactions in Table~\ref{table:reacs} are mainly hydrogenation reactions of common gas phase species and reactions between species that are likely to be abundant on the grains. Where possible, reactions that have been found to be efficient, or even dominant, routes to forming a species have been chosen. For example, the hydrogenation of CO to form CH$_3$OH is well studied \citep{Fuchs2009,Chuang2016} and so reactions 21-24 are the only considered route to form CH$_3$OH. Similarly, the formation of CO$_2$ via reaction 3 is known to be efficient \citep{Ioppolo2011} and other routes suffer from large energetic barriers. Note that in cases where H or H$_2$ is a product of a reaction such as in reaction 11, it is ignored and the total H abundance is not conserved. This is for simplicity and the lost H represents too small a fraction of the H abundance to affect the model.\par
There is no gas phase chemistry in the model and so the freeze out of species from the gas phase must be parameterized. The adsorption rate is assumed to be zero for all but the following six species: CO, CS, O, H, OH and S. To derive the adsorption rate of these species, the gas-grain chemical code \textsc{uclchem} \citep{Holdship2017} was utilized. \textsc{uclchem} was run with a network of 220 species with gas-phase reactions from UMIST12 \citep{mcElroy2013}, freeze out of gas phase species and the non-thermal desorption of grain surface species. A single point model of this full gas-grain chemistry was run in which the gas increased in density under freefall from \SI{e2}{\per\centi\metre\cubed} to \SI{2e4}{\per\centi\metre\cubed}, which is appropriate for a dark molecular cloud. The chemistry progresses over \SI{10}{\mega\year} at \SI{10}{\kelvin} and the freeze out rates for the six species above were extracted from this model.\par
The freeze out rates from \textsc{uclchem} were inserted as source terms in the ODEs for those species in the simple grain surface model. The grain surface models starts with an abundance of zero for all species, representing bare grains. The model then progresses for \SI{10}{\mega\year} considering only the 24 reactions in Table~\ref{table:reacs}, the freeze out rates and the non-thermal desorption of each species. In this way, the grain surfaces in a dark molecular cloud are effectively modelled whilst the computation time is low as the gas phase treatment is reduced to the incoming (freeze out) and outgoing (desorption) flux of molecules. Note that the cloud age is arbitrary, the model reaches the molecular cloud density at \SI{6}{\mega\year} and the chemistry is then allowed to progress until a total age of \SI{10}{\mega\year}. The exact choice of final time has only a small affect on the abundances in the model.\par
Whether one parameterizes the rate of surface reactions in a similar way to the Kooji-Arrhenius equation used for gas phase chemistry \citep{Occhiogrosso2012} or considers the diffusion and reaction of species across the ice surface \citep{Hasegawa1992,Chang2007}, the rate of a reaction is constant for a given temperature and dust composition. Therefore, $k$ in this model is treated as a constant rate of reaction in units of \si{\centi\metre\cubed\per\second} as the temperature in the model is constant at \SI{10}{\kelvin}. This reduces the number of parameters in the model and reflects the available data, i.e. ice phase abundances in quiescent, approximately isothermal clouds. \par
The result of these approximations and modifications is a model of the grain surface chemistry under the conditions of a dark molecular cloud at a constant temperature of \SI{10}{\kelvin}. The freeze out rates and gas density are specific to a dark cloud and so the model is not of applicable to arbitrary ices. This model has a run time that is approximately 1000 times shorter than an equivalent run of \textsc{uclchem}. This reduction in run time is vital due to the number of model runs required for an MCMC inference procedure.
\begin{table}[t]
\centering
\begin{tabular}{c}
\hline\noalign{\smallskip}
Species\\
\hline\noalign{\smallskip}
CH$_3$OH, CO, CO$_2$, CS, C$S_2$, H, H$_2$CO, H$_2$CS,\\
H$_2$O, H$_2$S,H$_2$$S_2$, HCO, HCS, HOCS, HS, HSO,\\ 
O, OCS, OH, S, SO, SO$_2$ \\
\hline
\end{tabular}
\caption{Species included in the chemical model.}
\label{table:species}
\end{table}
\begin{table}[t]
\centering
\begin{tabular}{c c c c c c}
\hline\noalign{\smallskip}
No. & \multicolumn{5}{c}{Reactions}\\
\hline\noalign{\smallskip}
1.&O &$+$  & H  &$\rightarrow$& OH \\ 
2.&OH &$+$  & H  &$\rightarrow$& H$_2$O \\
3.&CO &$+$  & OH  &$\rightarrow$& CO$_2$ \\
4.&S &$+$  & H  &$\rightarrow$& HS \\
5.&HS &$+$  & H  &$\rightarrow$& H$_2$S \\
6.&H$_2$S &$+$  & S  &$\rightarrow$& H$_2$$S_2$ \\
7.&CS &$+$  & H  &$\rightarrow$& HCS \\
8.&HCS &$+$  & H  &$\rightarrow$& H$_2$CS \\
9.&CO &$+$  & S &$\rightarrow$& OCS \\
10.&OCS &$+$  & H  &$\rightarrow$& HOCS \\
11.&H$_2$S &$+$  & CO &$\rightarrow$& OCS \\
12.&H$_2$S &$+$  & H$_2$S  &$\rightarrow$& H$_2$$S_2$ \\
13.&H$_2$$S_2$ &$+$  & CO &$\rightarrow$& CS2 $+$ O \\
14.&H$_2$S &$+$  & O  &$\rightarrow$& SO$_2$ \\
15.&C$S_2$ &$+$  & O  &$\rightarrow$& OCS $+$ S \\
16.&CO &$+$  & HS  &$\rightarrow$& OCS \\
17.&S &$+$  & O  &$\rightarrow$& SO \\
18.&SO &$+$  & O  &$\rightarrow$& SO$_2$ \\
19.&SO &$+$  & H  &$\rightarrow$& HSO \\
20.&HSO &$+$  & H  &$\rightarrow$& SO \\
21.&CO &$+$  & H  &$\rightarrow$& HCO \\
22.&HCO &$+$  & H  &$\rightarrow$& H$_2$CO \\
23.&H$_2$CO &$+$  & H  &$\rightarrow$& H$_3$CO \\
24.&H$_3$CO &$+$  & H  &$\rightarrow$& CH$_3$OH \\
\hline
\end{tabular}
\caption{Reaction Network used in the chemical model. The rates of these reactions are the parameters of interest in this work.}
\label{table:reacs}
\end{table}

\section{Bayesian Inference}
\label{sec:inference}
\subsection{Inference Procedure}
The aim of this work is to obtain information about the set of reaction rates $ \boldsymbol{k}=(k_1,k_2,...,k_{23} )$ of the surface chemical network, where $k_j$ is the reaction rate of reaction $j$. For a given set of rates, the model produces simulated molecular abundances $ \boldsymbol{\mathcal{Y}}=(\mathcal{Y}_1,\mathcal{Y}_2,...,\mathcal{Y}_{22} )$ where $\mathcal{Y}_{i}$ is the abundance of species $i$. These quantities are related through the chemical code $\mathcal{C}(\cdot)$, so that $ \boldsymbol{\mathcal{Y}}=\mathcal{C}(\boldsymbol{k})$.\par
For any set of simulated abundances,the probability of the corresponding parameter values can be evaluated through the use of Bayes' rule,
\begin{equation}
P(\boldsymbol{k}|\boldsymbol{d}) = \frac{P(\boldsymbol{d}|\boldsymbol{k})P(\boldsymbol{k})}{P(\boldsymbol{d})} \propto P(\boldsymbol{d}|\boldsymbol{k})P(\boldsymbol{k})
\end{equation}
where $\boldsymbol{d}$ is the data, representing a set of observational constraints on $\boldsymbol{\mathcal{Y}}$. $P(\boldsymbol{k}|\boldsymbol{d})$ is the posterior probability distribution (PPD) of $\boldsymbol{k}$ which expresses the level of certainty about the reaction rates after considering the data and any prior information. The denominator is known as the Bayesian evidence but for the purposes of this study can simply be treated as a normalization factor. The prior probability distribution ($P(\boldsymbol{k})$) adopted for the reaction rates is a logarithmically uniform distribution that is non-zero when the reaction rates are between $10^{-5}$ and $10^{-30}$ and zero elsewhere. The limits of the prior distributions represent a larger range than that of rates typical of gas phase reactions. This is a choice to reflect the exploratory nature of the work and is expected to cover all likely rates regardless of the nature of surface reactions.\par
The likelihood function $P(\boldsymbol{d}|\boldsymbol{k})$ must give the likelihood of having obtained the data given the assumed set of rates. Here the likelihood encodes measurement noise and is given as,
\begin{equation}
 P(\boldsymbol{d}|\boldsymbol{k}) = \exp\left(-\frac{1}{2}\sum_i{\frac{(d_i-\mathcal{Y}_i)^2}{\sigma_i^2}}\right)
 \label{eq:likelihood}
\end{equation}
where $\mathcal{Y}_i$ are the model abundances of each species for which there is data and $\sigma_i$ is the Gaussian uncertainty of each observed fractional abundance.\par
\subsection{Implementation and Data}
In order to constrain the reaction rates, data is required in the form of species abundances in the ices. The solid state fractional abundances of species in quiescent gas illuminated by background stars were taken from a comprehensive recent review \citep[][Table 2]{Boogert2015}. Of the species in the grain surface model, this review gives constraints on the abundance of H$_2$O, CO, CO$_2$ and CH$_3$OH. The value of the median fractional abundance is provided for each along with the upper and lower quartile values. To formulate the likelihood, it is assumed these values describe a Gaussian distribution for the abundance of each species. It is also assumed that the uncertainties on the abundances are independent which is likely given that the abundances and statistics presented by \citet{Boogert2015} are combinations of different data sets for each species. The median value is taken as the mean and the upper and lower quartile values can then be assumed to be 0.68$\sigma$ from that mean. The resulting abundances and uncertainties are listed in the upper half of Table~\ref{table:data}. Due to the low number of observations, these distributions are not perfect representations of the data as the quartile values are not precisely symmetric about the median.\par
\begin{table}
\centering
\begin{tabular}{cc}
\hline
Species & Abundance \\
\hline
H$_2$O & \num{4.0\pm1.3e-5} \\
CO & \num{1.2\pm0.8e-5} \\
CO$_2$ & \num{1.3\pm0.7e-5} \\
CH$_3$OH & \num{5.2\pm2.4e-6} \\
\hline 
H$_2$S & \textless \num{1.6e-6}\\
H$_2$CO & \textless \num{3.0e-5}\\
OCS & \textless \num{1.2e-7}\\
SO$_2$ & \textless \num{4.0e-6}\\
\hline
\end{tabular} 
\caption{Upper section: adopted abundances and uncertainties of species observed in the ices used as data in the parameter inference. Lower section: upper limits of the fractional abundance for other species which are used to further constrain the reactions rates. All values adapted from \citet{Boogert2015} as discussed in the text.}
\label{table:data}
\end{table}

In order to evaluate the posterior distribution function for all values of $\boldsymbol{k}$, a sampler must be used. The emcee python package \citep{Foreman-Mackey2013} was chosen for this purpose. This in an implementation of the affine-invariant monte carlo sampler proposed by \citet{Goodman2010}. Rates were sampled by 128 ``walkers'', each producing chains of $\sim$\num{e6} samples where the frequency of the appearance a particular rate value in the chain is proportional to its likelihood. These walkers start from random positions in rate space (ie all 24 rates have a random value from \SIrange{e-30}{e-5}{\centi\metre\cubed\per\second}).The sampling took approximately 100 hours using a single node on the DiRAC CSD3 platform's Skylake-Peta4 system utilizing emcee's built in MPI tools. In appendix~\ref{append:convergence}, heuristics are presented which demonstrated the chains are likely to have converged.\par
\subsection{Upper Limits}
The abundances of H$_2$O, CO, CO$_2$ and CH$_3$OH are the only strong constraints on the abundances of the species in this model. The reaction rates that are acquired as a result of performing a Bayesian parameter inference procedure with these abundances are presented and discussed in Section~\ref{sec:gaussresults}. However, weaker constraints do exist for other species. \citet{Boogert2015} provide upper limits on the abundances of OCS in dark clouds as well as upper limits on H$_2$CO, SO$_2$ and H$_2$S in other objects, the upper limits used are given in Table~\ref{table:data}. Therefore, a second parameter inference procedure was performed which was identical to the first except that the likelihood function was modified to take into account the upper limits. In order to be conservative, when deriving upper limits on species that have not been detected towards background stars, the upper limits towards YSOs were increased by an order of magnitude to allow for larger abundances in molecular clouds. This is the case for the upper limits on H$_2$CO and SO$_2$.\par
To account for these upper limits, modifications were made to Equation~\ref{eq:likelihood}. When including the upper limits, the likelihood of a model was calculated as,
\begin{equation}
 P(\boldsymbol{d}|\boldsymbol{k}) = \exp\left(-\frac{1}{2}\delta_i\sum_i{\frac{(d_i-\mathcal{Y}_i)^2}{\sigma_i^2}}\right) \left(1-S(C_i)\right)^{1-\delta_i}
 \label{eq:upperlikelihood}
\end{equation}
where $\delta_i$ is 1 for observed species and 0 for species with upper limits. $C$ is the upper-limit of a species' abundance and $S(C_i)$ is the survival function. This modification to the likelihood is standard for left-censored data, i.e. ones where a detection limit provides only an upper limit on a quantity \citep{Klein2003}. The survival function for a Gaussian distribution is used,
\begin{equation}
S(C_i) = 1 - \frac{1}{2}\left[1+erf\left(\frac{C_i-\mathcal{Y}_i}{\sigma_i}\right)\right]
\end{equation}
where $erf()$ is the error function. $\sigma_i$ is assumed to be a third of the value of the upper limit. \par
Equation~\ref{eq:upperlikelihood} is equivalent to Equation~\ref{eq:likelihood} for detected species. However, for upper limits it takes the value of $1-S$. For model abundances much less than the upper limit, this likelihood is equal to 1 and for model abundances much larger than the limit it takes a value of zero. Thus, whilst model abundances close to the upper limit are accepted, models with much larger predicted abundances have a likelihood of zero.
\subsection{Testing the Method}
\label{sec:methodtest}
In order to ascertain whether this method would be able to predict reactions rates from measured abundances in the case that the model was an accurate representation of reality, it was tested using abundances obtained from the model itself. First, random rates for all 24 reactions were generated and the model was run using those rates. The abundances of the four species in Table~\ref{table:data} for which observed abundances are available were then stored. ``Noisy'' abundances were then generated by drawing randomly from a Gaussian distribution with a mean value of the model abundances and a standard deviation set to a 50\% error. This produced four ``observations'' obtained from the model with known rates. The 50\% error was chosen as it is the approximate fractional error of the real observations. The MCMC procedure was then performed to see whether the known reaction rates could be obtained.\par
It was found that the majority of rates could not be recovered. However, in tests where the rates for reactions that produced H$_2$O, CO, CO$_2$ and CH$_3$OH were high enough to produce observable abundances, the rates of those reactions were recovered. That is to say that intrinsically, this method appears to only be able to give information on the rates of reactions that form the species for which observational constraints are available. 
\section{Results}
\label{sec:results}
\subsection{Gaussian Abundance Constraints}
\label{sec:gaussresults}
\begin{figure*}
\includegraphics[width=0.9\textwidth]{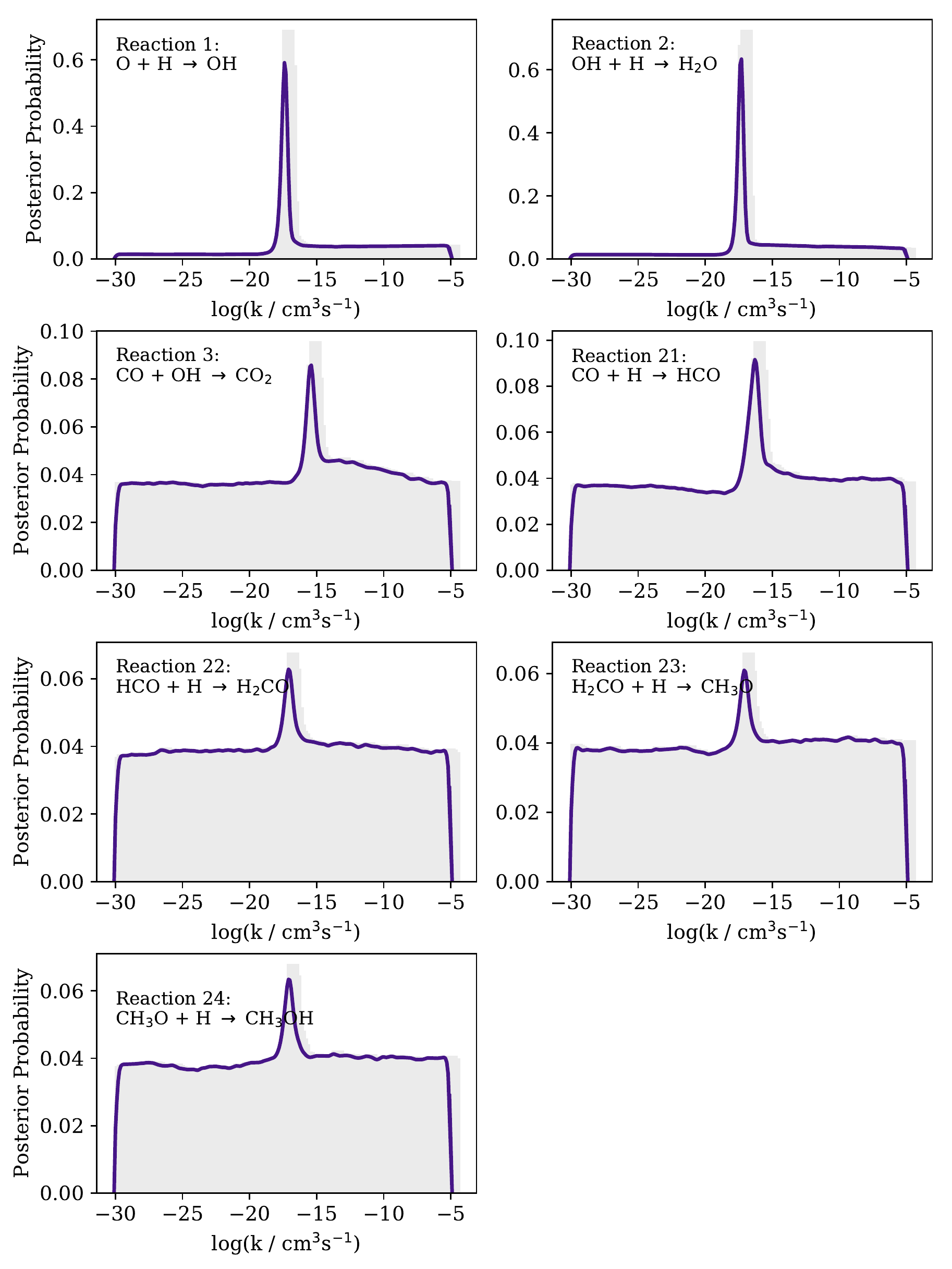}
\caption{The marginalised posterior probability distributions of the rates of the reactions with posterior distributions that are well constrained. The y-axis scaled such that the total probability density contained by the histogram is 1. All other reactions have posterior distributions that are approximately flat and similar to the prior. \label{fig:gaussianposteriors}}
\end{figure*}

The results are presented in the form of marginalized PPDs for the reaction rate coefficients. The density of each marginalized PPD reveals the areas where the corresponding reaction rate is more probable based on the imposed constraints. In Figure~\ref{fig:gaussianposteriors}, the marginalized PPDs of selected reactions are plotted. These show a large probability density only for a much smaller range of values than the prior. As expected from the tests in Section~\ref{sec:methodtest}, these are generally reactions that form the constrained species. The PPDs are shown as histograms and Gaussian kernel density estimates, using the full MCMC chains from all 128 walkers. It is believed that these chains have converged and the relevant tests are discussed in Appendix~\ref{append:convergence}.\par 
The PPD of reactions 1, 2, 3 and 21 are well constrained and involve species directly constrained by observation. Reaction 1 provides OH required to form H$_2$O and CO$_2$ through reactions 2 and 3. Those are in turn constrained by the observed abundances of CO$_2$ and H$_2$O and mutual competition for OH. Reactions 21 uses up CO and so it is expected there would be an upper limit due to the observed abundance of CO and a lower limit due to competition with reaction 3. Reactions 22-24 form CH$_3$OH from HCO. The competition between reactions and the correlation between the rates of reactions 21 through 24 is explored further in Section~\ref{sec:joints}.\par
The other PPDs are broadly similar to the prior distributions and the implications of this should be stated. Essentially, the reactions where the rates have uniform probability distributions are reactions that do not impact the likelihood of the model. It should be noted, however, that changes in the abundance of species not included in the likelihood calculation are possible. PPDs that are similar to the priors indicate that when modelling only H$_2$O, CO, CO$_2$ and CH$_3$OH, the rates of those reactions are unimportant. \par
\begin{table}
\centering
\begin{tabular}{ccc}
\hline
Reaction & Rate  & 65\% Probability Range\\
		 & (\si{\centi\metre\cubed\per\second}) & (\si{\centi\metre\cubed\per\second}) \\
\hline
1  & \num{4.0e-18} &  \num{1.0e-18} - \num{3.2e-10}\\
2  & \num{5.0e-18} &  \num{1.5e-18} - \num{1.5e-10}\\
3  & \num{4.0e-16} &  \num{6.8e-26} - \num{2.2e-10}\\
21  & \num{5.0e-17} & \num{6.6e-26} - \num{3.2e-10}\\
22  & \num{7.9e-18} & \num{4.6e-26} - \num{3.2e-10}\\
23  & \num{7.9e-18} & \num{4.6e-26} - \num{3.2e-10}\\
24  & \num{7.9e-18} & \num{4.6e-26} - \num{4.6e-10}\\
\hline
\end{tabular} 
\caption{Most likely values for the rates of well constrained reactions. The intervals containing 65\% of the probability density of the marginalized posteriors are also noted.}
\label{table:rates}
\end{table}
\subsection{Inclusion of Upper Limits}
\label{sec:upperlimits}
The PPDs of each reaction are largely unchanged when upper limits are included indicating that the upper limits may in fact be too conservative. The only major change is that the rate of reaction 10 takes a minimum value of \SI{e-17}{\centi\metre\cubed\per\second}. This is required for a models to produce a lower OCS abundance than the upper limit. \par
One may expect the upper limit on H$_2$CO to improve the level of certainty in the reaction series 21-24. However, using the conservative value in Table~\ref{table:data}, no change is seen in the posteriors. If the value for YSOs is utilized instead, only a small change is observed. In that case, the peaks that were apparent in the marginalized posterior of each reaction in Figure~\ref{fig:gaussianposteriors} become more pronounced, such that the majority of the probability density lies within them. Thus the value of H$_2$CO would represent an important constraint if an appropriate value for molecular clouds could be obtained.\par
The most likely rates for the well constrained reactions from this MCMC analysis are presented in Table~\ref{table:rates}. These most likely rates are the same whether they are taken from the observed abundances MCMC chains or the chains that included upper limits. It is clear that the available upper limits are not sufficiently constraining to improve the parameter inference.
\section{Discussion}
\label{sec:discussion}
\subsection{Model Abundances}
\label{sec:variance}
The results of the MCMC run give the marginalized posterior distributions of the rates of each reaction. In order to understand how well these rates reproduce the observed abundances, the model must be run with rates drawn from the probability distributions. This will allow the uncertainty in the model that arises from the uncertain rates to be quantified. The model was run 1000 times, with the rates of the reactions randomly sampled from the marginalized posterior distributions derived from the upper limit MCMC procedure. By plotting the average abundance of each species and the standard deviation of those abundances, the uncertainty in the model can be evaluated.\par
\begin{figure}[h]
\includegraphics[width=0.5\textwidth]{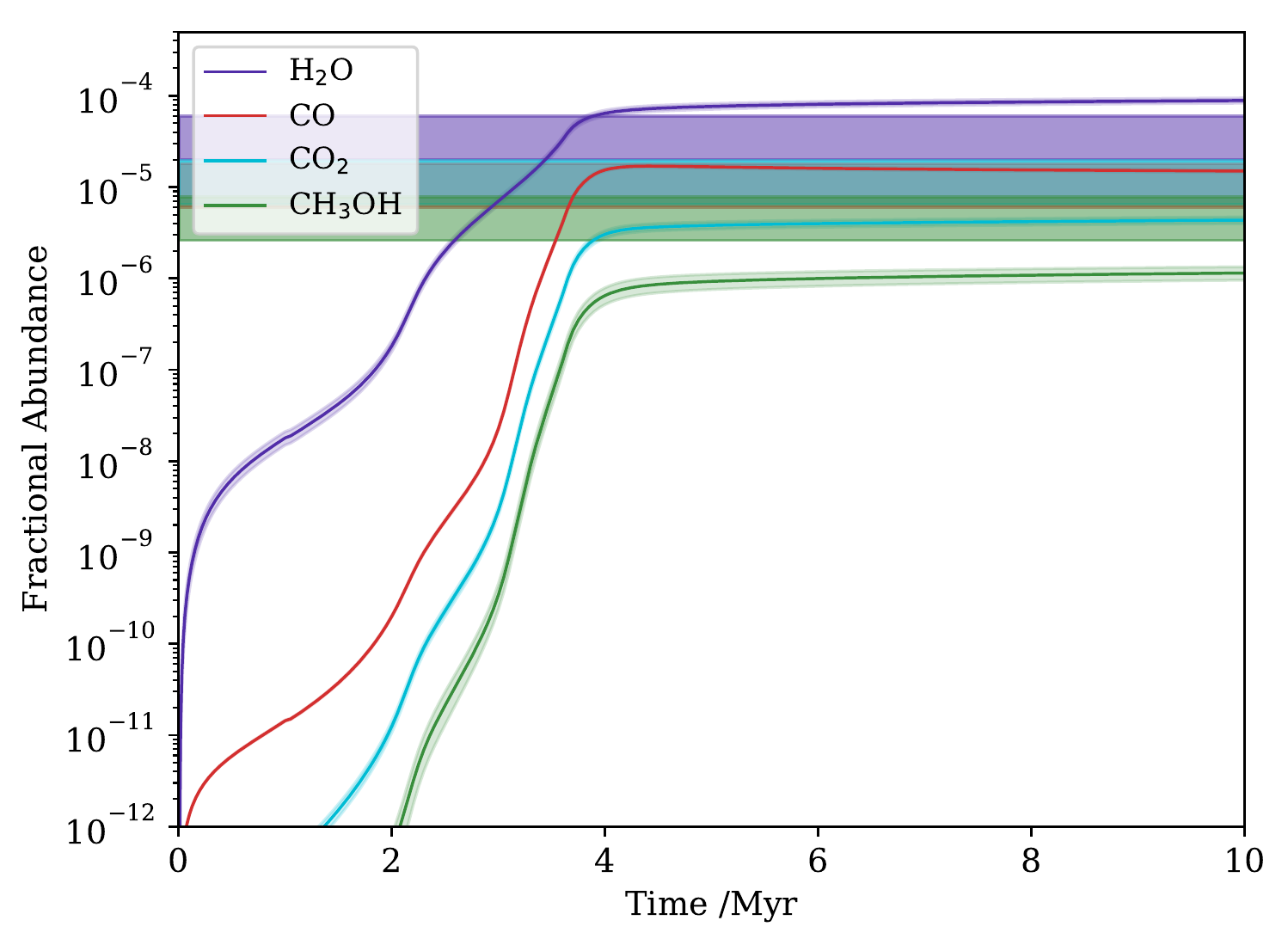}
\caption{Fractional abundances for the constrained species from 1000 model runs. The rates of each reaction are randomly sampled from the marginalized posterior distributions. The average abundances produced by the models at each time is plotted along with the 67\% confidence interval displayed as a shaded area.\label{fig:variance}}
\end{figure}
\begin{figure}[h]
\includegraphics[width=0.5\textwidth]{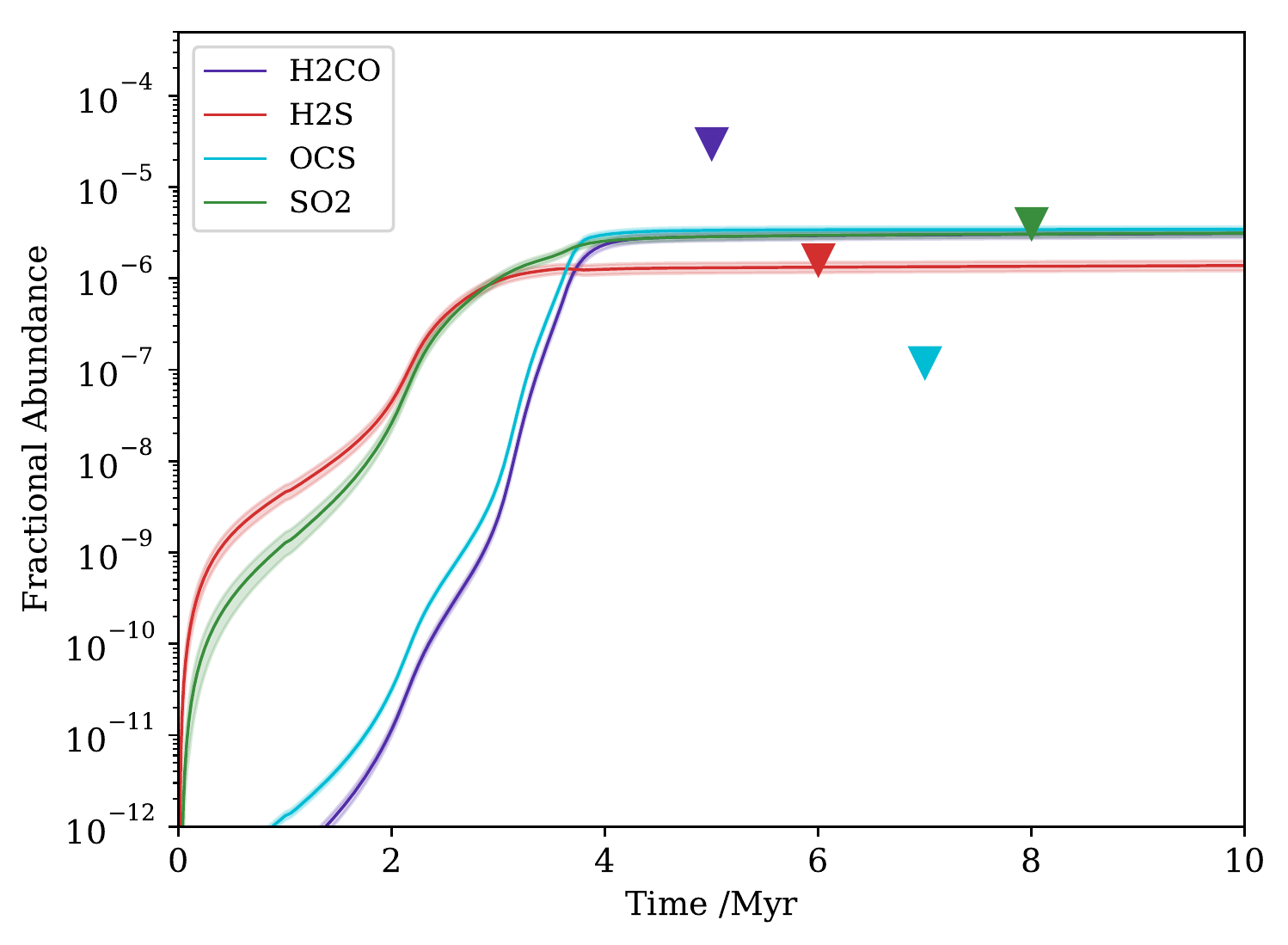}
\caption{Similar to Figure~\ref{fig:variance} but for the species with upper limits. The values of the upper limits are plotted as triangles. \label{fig:variance2}}
\end{figure}

The abundances for selected species from the model runs are plotted in Figures~\ref{fig:variance} and \ref{fig:variance2}. It can immediately be seen from Figure~\ref{fig:variance} that, for the species with observational constraints, the uncertainty in the rates does not lead to a large uncertainty in the model abundances. The model appears to consistently underproduce CH$_3$OH and overproduce H$_2$O. However, the difference is small if the errors on the observations are accounted for. This is a good result for such a simple model. It may be that if the reduced network was expanded, these results would be improved. Equally, it may be that the constraints are broad enough that a poor CH$_3$OH abundance is not affecting the overall likelihood as much as a poor CO or CO$_2$ abundance would.\par
The fractional abundances of the species with upper limits were not as well constrained and so it might be expected that they are much more varied. In Figure~\ref{fig:variance2} it can be seen that this is not the case, which implies that their abundances are also strongly tied to the rates of the reactions in Figure~\ref{fig:gaussianposteriors}. However, the average abundance of OCS is an order of magnitude higher than the observed upper limit. Examining the full abundance distribution, it appears there is a fraction of the model runs that fall outside the 67\% confidence interval and give OCS abundances below the upper limits. This illustrates the problem inherent in drawing from the marginalized posteriors. Drawing from each marginalized posterior individualy gives sets of reaction rates that break the upper limit used to infer the posterior distributions.\par
\subsection{Network Connectivity}
\label{sec:joints}
The PPDs presented in Figure~\ref{fig:gaussianposteriors} are marginalized, that is to say they represent the likelihood of a given reaction rate averaged over the values of the other rates. However, not all rates are independent and it is possible that some areas of the rate space are only likely for one reaction when a second takes a particular value. To investigate this, the joint posteriors of pairs of reactions were examined. These give the likelihood of pairs of reactions rate values so that it can be seen whether the two reactions are in some way correlated.\par
\begin{figure}
\includegraphics[width=0.5\textwidth]{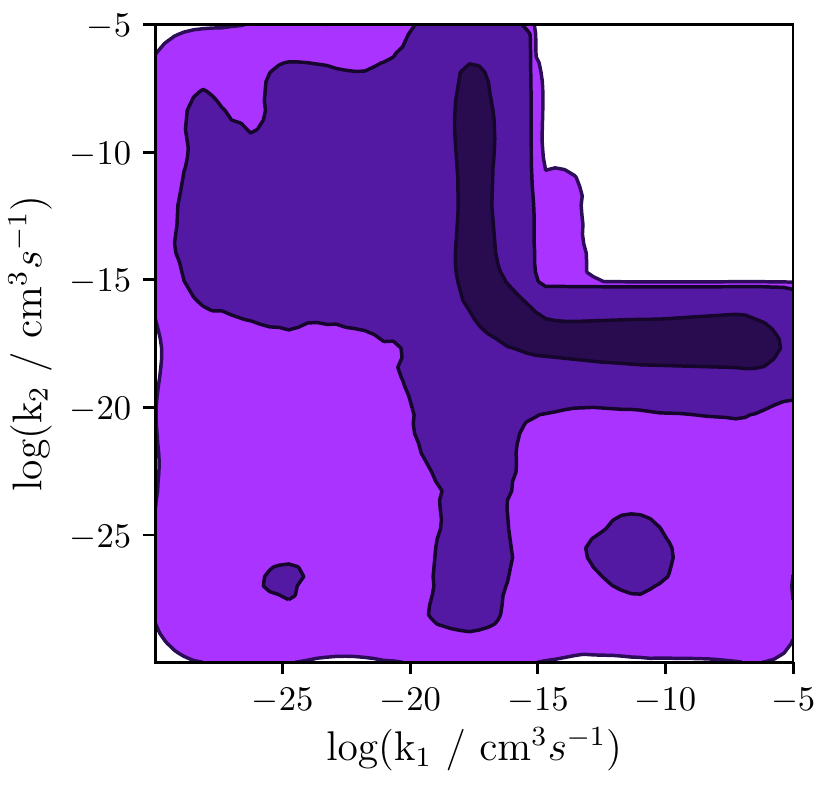}
\caption{The joint probability distributions of the rates of reactions 1 and 2, darker areas represent higher probability densities. The 1,2 and 3 sigma contours are plotted. These two reactions are tightly coupled, one must take the value of $\sim$\SI{5.0e-18}{\centi\metre\cubed\per\second} and act as the rate limiting step. The other must then have this value or higher in order for enough H$_2$O to be produced in the model. \label{fig:joint1-2}}
\end{figure}
For example, the joint probability distribution of reactions 1 and 2 is shown in Figure~\ref{fig:joint1-2}. It can be seen that either reaction 1 or reaction 2 can take a value much higher than their respective most likely value but only when the other is at its most likely value. This shows that, in reality, there must be certain amount of O converted to H$_2$O in the model and as long as one step in that process limits the rate to the correct amount then the other can freely vary. In order to break this degeneracy, limits on the OH abundance in the ice are required. Similar joint distributions are seen for reactions 22 to 24, as a certain amount of CO must be converted to CH3OH.\par
\begin{figure}
\includegraphics[width=0.5\textwidth]{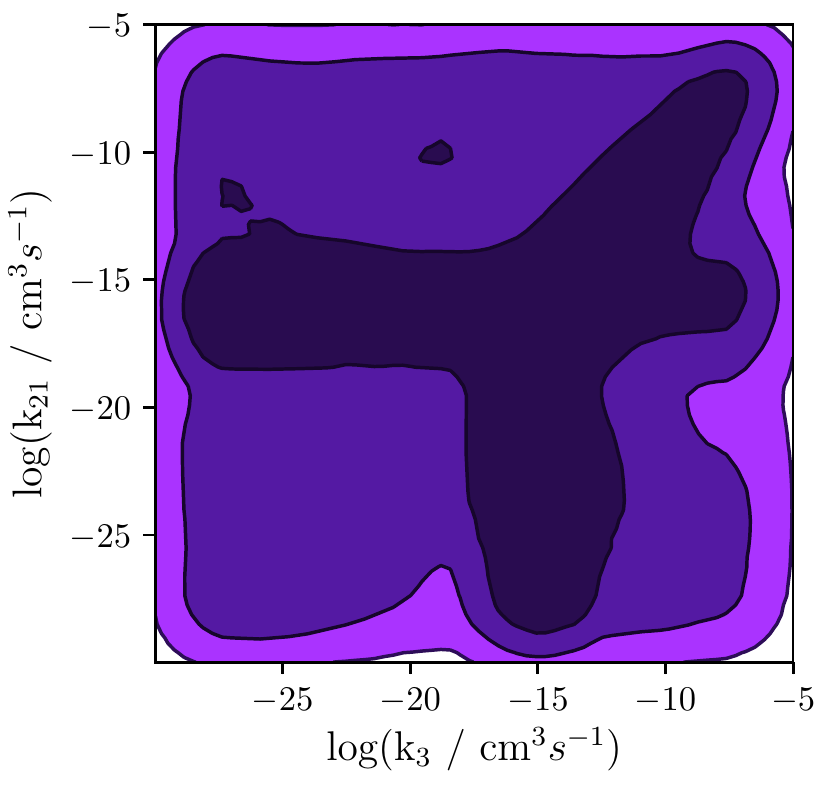}
\caption{Similar to Figure~\ref{fig:joint1-2} for reactions 3 and 21. Whilst a large amount of the probability density is at the location of the peaks in the marginalized posterior distributions, there is a noticeable correlation when either increases above the most likely rate. This is likely due to the fact that both reactions destroy CO and are therefore competing to produce enough CO$_2$ and HCO respectively. \label{fig:joint3-21}}
\end{figure}
In Figure~\ref{fig:joint3-21}, the joint probability distribution of reactions 3 and 21 are plotted. Both reactions are less well constrained than reactions 1 or 2. This can be seen from the large area taken up by the 1$\sigma$ contour. However, the high probability density areas are those where at least one reaction takes the most likely value from their respective marginalized posterior distributions. There is also a line of increased probability density where both reactions have approximately equal rates that are higher than the peak value. It is therefore likely that the reactions compete for CO and the rate of reaction 3 is poorly constrained as the availability of CO is the main factor in the amount of CO$_2$ produced. Tighter observational ranges on the abundances of CO, CO$_2$ and CH$_3$OH may reduce this degeneracy and allow the rate of reaction 3 to be more clearly determined.
\section{Conclusions}
\label{sec:conclusions}
A novel way to tackle uncertainty about surface reactions and rate coefficients using Bayesian inference was presented. To prove the efficiency of Bayesian techniques in providing insight on the chemical parameters of surface reactions, the algorithm was tested with a proof of concept example. A simple chemical code was created by parameterizing the freeze out of important species and neglecting all other adsorption to the grain surface. This left a model where only grain surface chemistry needed to be accounted for, greatly reducing the complexity and opening up the possibility of exploring a large parameter space.\par
The rates of the reactions in the chemical model were found through Bayesian inference. Using an MCMC sampling algorithm, the model was run with varying reaction rates and the likelihood of the model was evaluated each time. This likelihood was calculated by comparing the model to observations of ices towards background stars, which are reasonable values for a molecular cloud. It was possible to strongly constrain the rates of reactions that are involved in the production or destruction of species for which measurements exist. These rates are presented in Table~\ref{table:rates}.\par
Future improvements should include a more complex chemical code, including the grain surface reactions directly in a gas-grain chemical code. This would allow an improved treatment of the freeze out and non-thermal desorption amongst other effects. However, the added complexity would make this a vastly more computationally intensive procedure, initial tests with \textsc{uclchem} taking approximately 1000 times longer per run. Improved rates could also be achieved by including more observational data, particularly constraining species for which there are currently only upper limits. The parameter space could also be reduced by including the PPDs from the results of this work as priors in future work.\par
\acknowledgments
The authors thank the referees for their constructive comments which greatly improved this work. They also thank Damien De Mijolla for conversations relating to the likelihood function. J.H. is funded by the STFC grant ST/M001334/1. A.M. was partially supported by an IMPACT studentship. The authors also acknowledge DiRAC for use of their HPC system which allowed this work to be performed. 

\appendix
\section{Convergence}
\label{append:convergence}
It is important to understand whether an MCMC procedure has converged to a stationary distribution. The sampling is not complete if increasing the length of that chains would considerably alter the posterior distribution of the reaction rates. It is not possible to be certain that convergence has been reached but several heuristics are available and are considered in this appendix.\par
\begin{figure}
\includegraphics[width=0.9\textwidth]{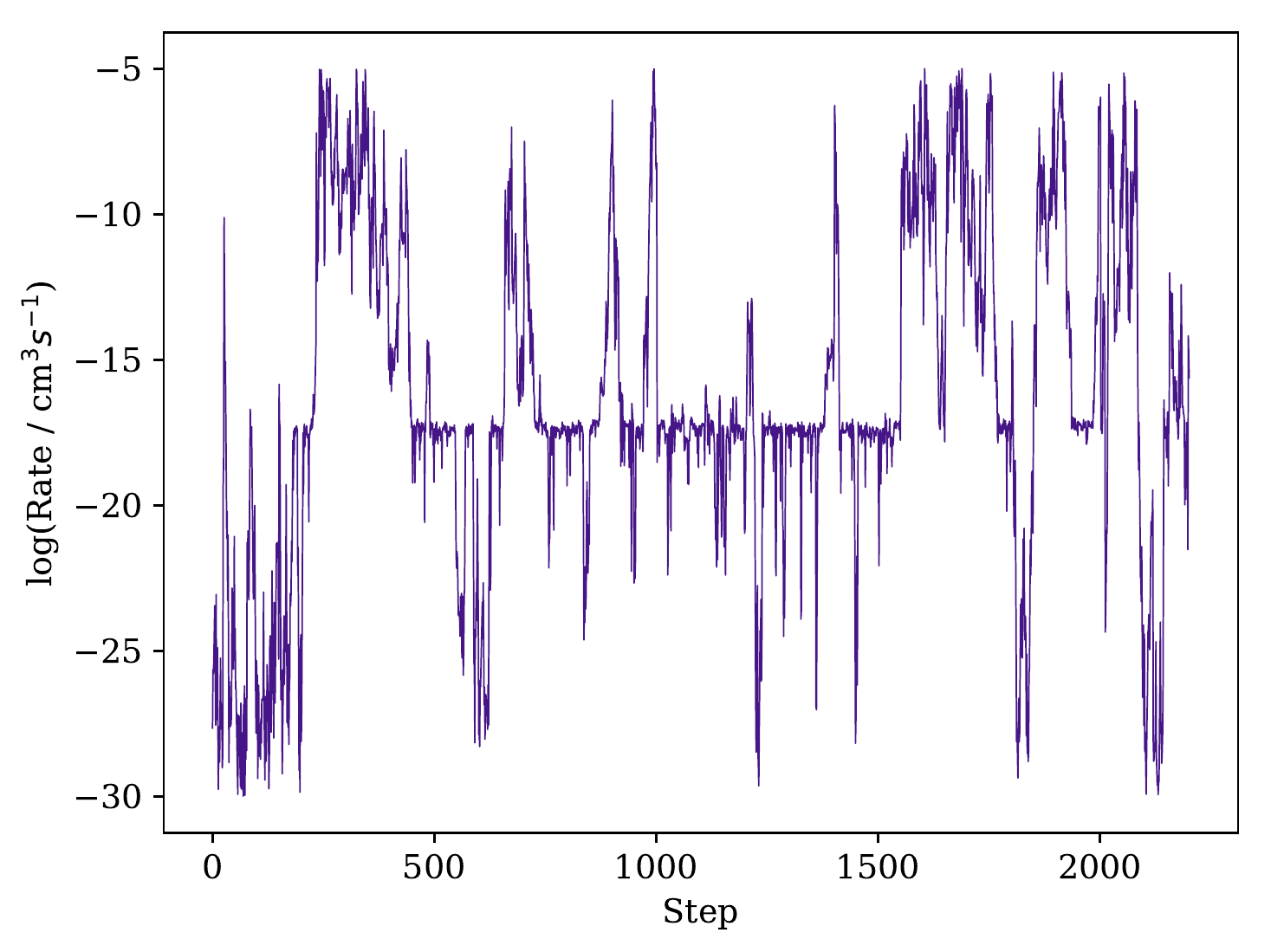}
\caption{Trace plots showing the sampled rate of reaction 1 for every 500th step in an example chain. The chain repeatedly returns to the optimal rate whilst fully exploring the range of possible rates.}
\label{fig:traceplot}
\end{figure}

Most simply, the chains themselves can be inspected. Figure~\ref{fig:traceplot} shows every 500th step in an example chain for reaction 1. The walker repeatedly leaves the area of maximum likelihood and then returns. This cycle repeats a sufficient number of times for it to be unlikely that there is a undiscovered mode.\par
\begin{figure}
\includegraphics[width=0.9\textwidth]{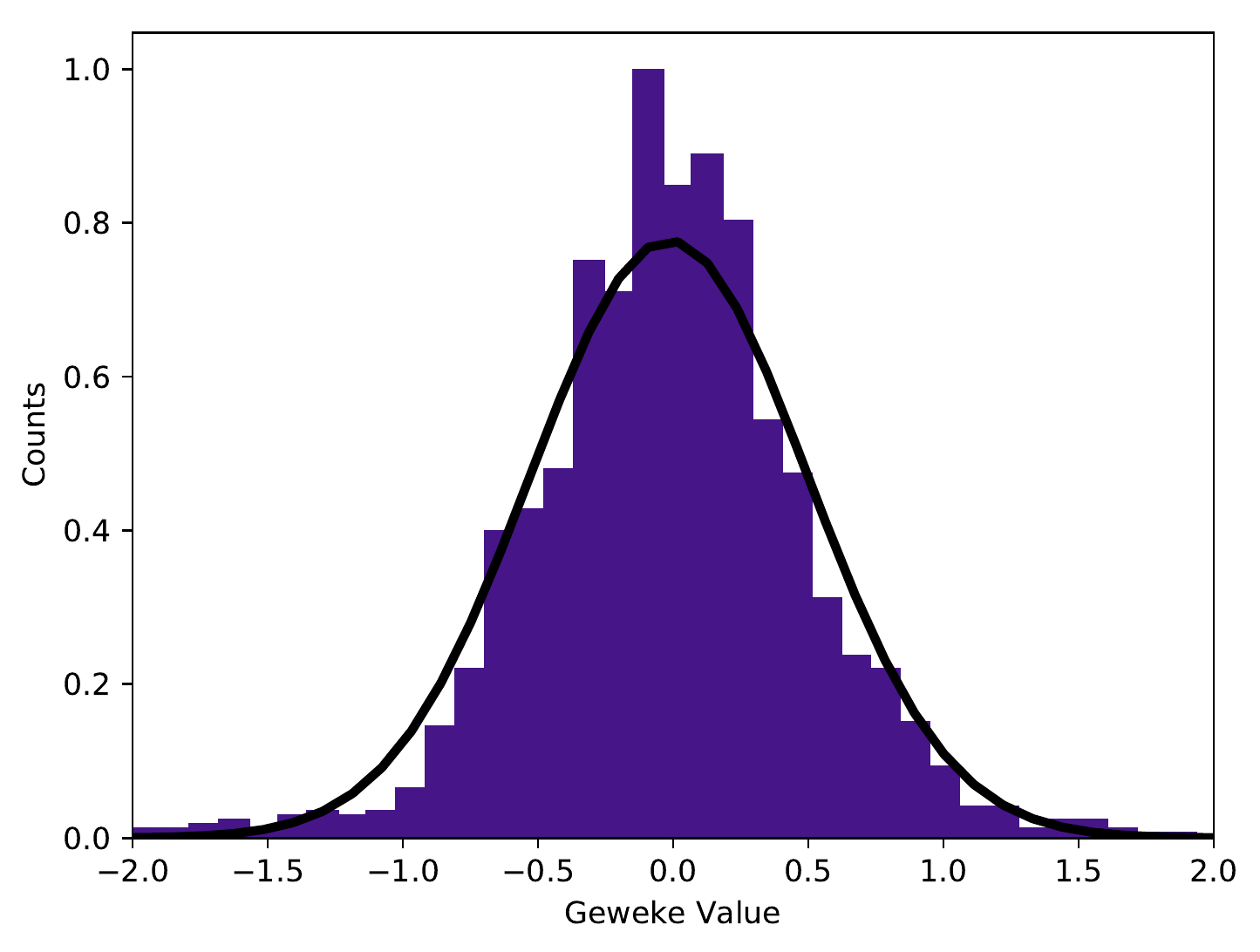}
\caption{Geweke diagnostic for subsamples of the MCMC chains run for this work. The distribution of the values of the diagnostic should follow a normal distribution if the chains have converged. A normal distribution with the same standard deviation as the Geweke diagnostic samples has been plotted in black for comparison. }
\label{fig:geweke}
\end{figure}

More rigorously, the Geweke diagnostic can be used \citep{Geweke1992}. In this test, it is considered that if the chain has converged any two samples of the chain will have the same mean, within the variance of the samples. This is typically tested on the first 10\% and the final 50\% of a chain that is thought to have converged. In this work, a sample of chains were tested by breaking each chain into 10 subsamples and comparing each to the mean of the final 50\% of the chain. In every case, the mean of the subsample was consistent with the mean of the larger sample. Since the value of the Geweke diagnostic should be zero, within the variance of the chains, multiple values for converged chains should follow a normal distribution. In Figure~\ref{fig:geweke} the values of the diagnostic for many subsamples of the chains from this work are plotted as histogram with a normal distribution plotted for comparison. \par
The auto-correlation time is another diagnostic that can be calculated and the tutorial provided in the emcee documentation\footnote{\url{https://emcee.readthedocs.io/en/latest/tutorials/autocorr/}} is used for this. This provides two heuristics. Firstly, once the chains reach a sufficient length that the autocorrelation time can be reliably calculated, it is likely that the chain has converged. Secondly, one use of the quantity is to calculate the sampling error in an MCMC chain. An autocorrelation time of 10$^4$ steps was calculated from the chains, effectively giving approximately 100 independent samples per chain. If the mean value of the chain is considered then the variance on this mean is given through the equation,
\begin{equation}
\sigma^2 = \frac{\tau_f}{N}Var[f(\theta)]
\end{equation}
where N is the number of sample and $Var[f(\theta)]$ is the variance of the chain. In the case of an average chain in this work, the sampling error on the value of the mean is $\sim$1\%.\par
Finally, the posterior distribution was also evaluated using the code pyMultinest \citep{Buchner2014} and the marginalised posteriors were consistent with those found using emcee. The consistency between these two different methods of sampling the posterior is good evidence for convergence. Ultimately, given the above heuristics and the fact that the initial MCMC runs produced approximately the same results for chains of 100,000 steps as they do in the $\sim$\num{e6} step chains used in the final work, it is assumed that the chains have, in fact, converged.

\bibliography{grainsensitivity}

\begin{thebibliography}{}
\expandafter\ifx\csname natexlab\endcsname\relax\def\natexlab#1{#1}\fi

\bibitem[{Bevan(2018)}]{Bevan2018}
Bevan, A. 2018, Monthly Notices of the Royal Astronomical Society, 480, 4659

\bibitem[{Boogert {et~al.}(2015)Boogert, Gerakines, \& Whittet}]{Boogert2015}
Boogert, A. C.~A., Gerakines, P.~A., \& Whittet, D. C.~B. 2015, Annu. Rev.
  Astron. Astrophys, 53, 541

\bibitem[{Buchner {et~al.}(2014)Buchner, Georgakakis, Nandra, Hsu, Rangel,
  Brightman, Merloni, Salvato, Donley, \& Kocevski}]{Buchner2014}
Buchner, J., Georgakakis, a., Nandra, K., {et~al.} 2014, Astronomy {\&}
  Astrophysics, 564, A125

\bibitem[{Caselli \& Ceccarelli(2012)}]{Caselli2012}
Caselli, P., \& Ceccarelli, C. 2012, The Astronomy and Astrophysics Review, 20,
  56

\bibitem[{Ceccarelli {et~al.}(2007)Ceccarelli, Caselli, Herbst, Tielens, \&
  Caux}]{Ceccarelli2007}
Ceccarelli, C., Caselli, P., Herbst, E., Tielens, A. G. G.~M., \& Caux, E.
  2007, Protostars and Planets V, 47

\bibitem[{Chang {et~al.}(2007)Chang, Cuppen, \& Herbst}]{Chang2007}
Chang, Q., Cuppen, H.~M., \& Herbst, E. 2007, Astronomy {\&} Astrophysics, 469,
  973

\bibitem[{Chuang {et~al.}(2016)Chuang, Fedoseev, Ioppolo, van Dishoeck, \&
  Linnartz}]{Chuang2016}
Chuang, K.-J., Fedoseev, G., Ioppolo, S., van Dishoeck, E., \& Linnartz, H.
  2016, Monthly Notices of the Royal Astronomical Society, 455, 1702

\bibitem[{Collings \& McCoustra(2006)}]{Collings2006}
Collings, M.~P., \& McCoustra, M. R.~S. 2006, Proceedings of the International
  Astronomical Union, 1, 405

\bibitem[{Draine(2003)}]{Draine2003}
Draine, B. 2003, Annual Review of Astronomy and Astrophysics, 41, 241

\bibitem[{Foreman-Mackey {et~al.}(2013)Foreman-Mackey, Hogg, Lang, \&
  Goodman}]{Foreman-Mackey2013}
Foreman-Mackey, D., Hogg, D.~W., Lang, D., \& Goodman, J. 2013, Publications of
  the Astronomical Society of The Pacific, 306

\bibitem[{Fraser {et~al.}(2004)Fraser, Collings, Dever, \&
  McCoustra}]{Fraser2004}
Fraser, H.~J., Collings, M.~P., Dever, J.~W., \& McCoustra, M. R.~S. 2004,
  Monthly Notices of the Royal Astronomical Society, 353, 59

\bibitem[{Fuchs {et~al.}(2009)Fuchs, Cuppen, Ioppolo, Romanzin, Bisschop,
  Andersson, {Van Dishoeck}, Linnartz, \& Beverly}]{Fuchs2009}
Fuchs, G.~W., Cuppen, H.~M., Ioppolo, S., {et~al.} 2009, A{\&}A, 505, 629

\bibitem[{Geweke(1992)}]{Geweke1992}
Geweke, J. 1992, {Evaluating the Accuracy of Sampling-Based Approaches to the
  Calculations of Posterior Moments}, Vol.~4 (Clarendon Press), 641--649

\bibitem[{Goodman \& Weare(2010)}]{Goodman2010}
Goodman, J., \& Weare, J. 2010, Communications in Applied Mathematics and
  Computational Science, 5, 65

\bibitem[{Hagen {et~al.}(1979)Hagen, Allamandola, \& Greenberg}]{Hagen1979}
Hagen, W., Allamandola, L.~J., \& Greenberg, J.~M. 1979, Astrophysics and Space
  Science, 65, 215

\bibitem[{Hasegawa {et~al.}(1992)Hasegawa, Herbst, \& Leung}]{Hasegawa1992}
Hasegawa, T.~I., Herbst, E., \& Leung, C.~M. 1992, The Astrophysical Journal
  Supplement Series, 82, 167

\bibitem[{Herbst \& {Van Dishoeck}(2009)}]{Herbst2009}
Herbst, E., \& {Van Dishoeck}, E.~F. 2009, Annual Review of Astronomy and
  Astrophysics, 47, 427

\bibitem[{Holdship {et~al.}(2017)Holdship, Viti, Jim{\'{e}}nez-Serra,
  Makrymallis, \& Priestley}]{Holdship2017}
Holdship, J., Viti, S., Jim{\'{e}}nez-Serra, I., Makrymallis, A., \& Priestley,
  F. 2017, The Astronomical Journal, 154, 38

\bibitem[{Ioppolo {et~al.}(2009)Ioppolo, Palumbo, Baratta, \&
  Mennella}]{Ioppolo2009}
Ioppolo, S., Palumbo, M.~E., Baratta, G.~A., \& Mennella, V. 2009, Astronomy
  {\&} Astrophysics, 493, 1017

\bibitem[{Ioppolo {et~al.}(2011)Ioppolo, van Boheemen, Cuppen, van Dishoeck, \&
  Linnartz}]{Ioppolo2011}
Ioppolo, S., van Boheemen, Y., Cuppen, H.~M., van Dishoeck, E.~F., \& Linnartz,
  H. 2011, Monthly Notices of the Royal Astronomical Society, 413, 2281

\bibitem[{Klein \& Moeschberger(2003)}]{Klein2003}
Klein, J.~P., \& Moeschberger, M.~L. 2003, {Survival analysis : techniques for
  censored and truncated data} (Springer), 536

\bibitem[{Lomax {et~al.}(2016)Lomax, Whitworth, \& Hubber}]{Lomax2016}
Lomax, O., Whitworth, A.~P., \& Hubber, D.~A. 2016, Monthly Notices of the
  Royal Astronomical Society, 458, 1242

\bibitem[{Makrymallis \& Viti(2014)}]{Makrymallis2014}
Makrymallis, A., \& Viti, S. 2014, The Astrophysical Journal, 794, 45

\bibitem[{McElroy {et~al.}(2013)McElroy, Walsh, Markwick, Cordiner, Smith, \&
  Millar}]{mcElroy2013}
McElroy, D., Walsh, C., Markwick, A.~J., {et~al.} 2013, Astronomy {\&}
  Astrophysics, 550, A36

\bibitem[{Minissale {et~al.}(2015)Minissale, Loison, Baouche, Chaabouni,
  Congiu, \& Dulieu}]{Minissale2015}
Minissale, M., Loison, J.-C., Baouche, S., {et~al.} 2015, Astronomy {\&}
  Astrophysics, 577, A2

\bibitem[{Occhiogrosso {et~al.}(2012)Occhiogrosso, Viti, Ward, \&
  Price}]{Occhiogrosso2012}
Occhiogrosso, A., Viti, S., Ward, M.~D., \& Price, S.~D. 2012, Monthly Notices
  of the Royal Astronomical Society, 427, 2450

\bibitem[{Palau {et~al.}(2014)Palau, Zapata, Rodr{\'{i}}guez, Bouy, Barrado,
  Morales-Calder{\'{o}}n, Myers, Chapman, Ju{\'{a}}rez, \& Li}]{Palau2014}
Palau, A., Zapata, L.~A., Rodr{\'{i}}guez, L.~F., {et~al.} 2014, Monthly
  Notices of the Royal Astronomical Society, 444, 833

\bibitem[{Parise {et~al.}(2005)Parise, Ceccarelli, \& Maret}]{Parise2005}
Parise, B., Ceccarelli, C., \& Maret, S. 2005, Astronomy {\&} Astrophysics,
  441, 171

\bibitem[{Pirronello {et~al.}(1982)Pirronello, Brown, Lanzerotti, Marcantonio,
  \& Simmons}]{Pirronello1982}
Pirronello, V., Brown, W.~L., Lanzerotti, L.~J., Marcantonio, K.~J., \&
  Simmons, E.~H. 1982, The Astrophysical Journal, 262, 636

\bibitem[{Pirronello {et~al.}(1997)Pirronello, Liu, Shen, \&
  Vidali}]{Pirronello1997}
Pirronello, V., Liu, C., Shen, L., \& Vidali, G. 1997, The Astrophysical
  Journal, 475, L69

\bibitem[{Prager {et~al.}(2013)Prager, Najm, \& Z{\'{a}}dor}]{Prager2013}
Prager, J., Najm, H.~N., \& Z{\'{a}}dor, J. 2013, Proceedings of the Combustion
  Institute, 34, 583

\bibitem[{Schmalzl {et~al.}(2014)Schmalzl, Visser, Walsh, Albertsson, {Van
  Dishoeck}, Kristensen, \& Mottram}]{Schmalzl2014}
Schmalzl, M., Visser, R., Walsh, C., {et~al.} 2014, Astronomy {\&}
  Astrophysics, 572, A81

\bibitem[{Testi {et~al.}(2016)Testi, Natta, Scholz, Tazzari, Ricci, \& {de
  Gregorio Monsalvo}}]{Testi2016}
Testi, L., Natta, A., Scholz, A., {et~al.} 2016, Astronomy {\&} Astrophysics,
  593, A111

\bibitem[{van Dishoeck(2004)}]{Vandishoeck2004}
van Dishoeck, E.~F. 2004, Annual Review of Astronomy and Astrophysics, 42, 119

\bibitem[{Watanabe {et~al.}(2005)Watanabe, Nagaoka, Hidaka, \&
  Kouchi}]{Watanabe2005}
Watanabe, N., Nagaoka, A., Hidaka, H., \& Kouchi, A. 2005, Protostars and
  Planets V Posters, 8244

\bibitem[{Williams \& Cecchi-Pestellini(2015)}]{Williams2015}
Williams, D.~A., \& Cecchi-Pestellini, C. 2015, {The Chemistry of Cosmic Dust}
  (Royal Society of Chemistry)

\end{thebibliography}
\end{document}